\documentclass[aip, preprint]{revtex4-2}

\usepackage{amsmath,array}
\usepackage[version=4]{mhchem}
\usepackage{mathtools}
\usepackage{setspace}
\usepackage{xcolor}
\usepackage[utf8]{inputenc}
\usepackage{graphicx}
\usepackage{hyperref}
\hypersetup{
    colorlinks,
    filecolor=magenta,
    linkcolor=blue,
    citecolor=blue,      
    urlcolor=green!50!black
}

\begin{document}
\setcounter{page}{1}

\title{A Geometrical Approach to Neutrino Oscillation Parameters}

\author{Mohammad Ful Hossain Seikh}
\email{fulhossain@ku.edu}
\affiliation{Department of Physics and Astronomy, University of Kansas, Lawrence, KS, USA 66045}

\begin{abstract}
We propose a geometric hypothesis for neutrino mixing: that twice the sum of the three mixing angles equals $180^\circ$, forming a Euclidean triangle. This condition leads to a predictive relation among the mixing angles and, through trigonometric constraints, enables reconstruction of the mass-squared splittings. The hypothesis offers a phenomenological resolution to the $\theta_{23}$ octant ambiguity, reproduces the known mass hierarchy patterns, and suggests a normalized geometric structure underlying the PMNS mixing. We show that while an order-of-magnitude scale mismatch remains (the absolute splittings are underestimated by $\sim 10 \times$), the triangle reproduces mixing ratios with notable accuracy, hinting at deeper structural or symmetry-based origins. We emphasize that the triangle relation is advanced as an empirical, phenomenological organizing principle rather than a result derived from a specific underlying symmetry or dynamics. It is testable and falsifiable: current global-fit values already lie close to satisfying the condition, and improved precision will confirm or refute it. We also outline and implement a simple $\chi^2$ consistency check against global-fit inputs to quantify agreement within present uncertainties.
\end{abstract}

\maketitle


\section{Introduction}\label{sec:I}

Neutrino oscillation experiments have firmly established that neutrinos are massive and exhibit flavor mixing, parametrized by the ``Pontecorvo-Maki-Nakagawa-Sakata (PMNS) matrix" with three mixing angles ($\theta_{12}, \theta_{13}, \theta_{23}$), two mass-squared differences ($\Delta m^2_{21}, \Delta m^2_{3\ell}$), and a CP-violating phase ($\delta_{CP}$)~\cite{Kajita2016,PDG2024,Esteban2024,Seikh2020SSVP}. Despite impressive experimental precision, notable ambiguities remain; most prominently, the octant ambiguity of $\theta_{23}$, i.e., whether $\theta_{23}<45^\circ$ or $\theta_{23}>45^\circ$, and the ordering of the neutrino mass spectrum (normal vs inverted)~\cite{deSalas2020,Agarwalla2013}. Here, we follow the NuFIT convention where $\Delta m^2_{3\ell}$ denotes $\Delta m^2_{31}$ for normal ordering (NO) and $\Delta m^2_{32}$ for inverted ordering (IO), with $\ell$ indicating the lightest neutrino mass eigenstate. Using the latest NuFIT~6.0 2024 \cite{Esteban2024} and PDG~2024 \cite{PDG2024} global fit values for NO, we adopt,

\begin{equation}
\begin{aligned}
\theta_{12} &= 33.41^{+0.75}_{-0.72}{}^\circ, \quad
\theta_{13} = 8.54^{+0.11}_{-0.12}{}^\circ,
\quad
\theta_{23} = 49.1^{+1.0}_{-1.3}{}^\circ, \quad
\delta_{CP} = 197^{+42}_{-25}{}^\circ,\\
\Delta m^2_{21} &= (7.41^{+0.21}_{-0.20})\times10^{-5}\,\mathrm{eV}^2,
\quad
\Delta m^2_{3\ell} = (2.507^{+0.027}_{-0.026})\times10^{-3}\,\mathrm{eV}^2.
\end{aligned}
\label{eq:params}
\end{equation}

These values reflect improved sensitivity from long-baseline, reactor, and atmospheric neutrino experiments~\cite{Esteban2024,PDG2024}. Nonetheless, the octant ambiguity of $\theta_{23}$ persists; current fits slightly favor the second octant ($\theta_{23}>45^\circ$) but cannot exclude the first octant~\cite{deSalas2020}. Equally unresolved is whether the mass ordering is normal or inverted, although cosmological and oscillation data weakly prefer NO~\cite{deSalas2020,Esteban2024}.

In this work we treat the triangle sum as an empirical organizing principle for the observed PMNS parameters, rather than a consequence of a specific symmetry. With the representative central values in Eq.~\eqref{eq:params}, one finds $\theta_{12}+\theta_{13}+\theta_{23}\approx 91.05^\circ$, i.e., within $\sim\!1^\circ$ of $90^\circ$ (well inside current $1\sigma$ uncertainties on the angles), motivating a quantitative test of this correlation using present data~\cite{Esteban2024,PDG2024}. We introduce a novel geometrical interpretation; we hypothesize that twice the sum of the three mixing angles forms the internal angles of a triangle, exactly summing to $180^\circ$. Mathematically,
\begin{equation}
2(\theta_{12} + \theta_{13} + \theta_{23}) = 180^\circ.
\label{eq:triangle}
\end{equation}

This hypothesis is not derived from first principles but is instead motivated by the observation that current global-fit values of the mixing angles nearly satisfy this triangle condition. Such geometric structures are not unprecedented in particle physics; for example, the unitarity triangle in quark mixing arises from CKM matrix constraints. Within this triangle framework, the value of $\theta_{23}$ is geometrically determined by the other two mixing angles, providing a natural resolution to the octant ambiguity. Deviations from this triangle condition can point toward alternative mass-ordering scenarios. 
In particular, the triangle geometry yields not only normal and inverted mass hierarchies, but also a third formal possibility, a so-called \emph{mixed} hierarchy. We emphasize at the outset that this case is not physically distinct: it is equivalent to NO or IO under relabeling, and is included only as a formal outcome for completeness.

Throughout, we assume the standard three-neutrino, unitary PMNS framework and PDG parameterization~\cite{PDG2024}. The triangle relation reduces one free mixing parameter, thereby increasing predictivity: precise measurements of any two angles determine the third. This makes the hypothesis falsifiable, e.g., a confirmed first-octant value of $\theta_{23}$ or a precise sum significantly different from $90^\circ$ would rule out the exact relation.

In this work, we test the predictive power of the triangle hypothesis against the latest global-fit oscillation parameters, explore its implications for mass splittings and ordering, and investigate its compatibility with theoretical frameworks such as Tri-Bimaximal (TBM) and Bi-Maximal (BM) mixing, as well as symmetry-based models. We additionally implement a simple $\chi^2$ consistency check using NuFIT/PDG central values and $1\sigma$ uncertainties to quantify present-day agreement of the angle correlation. The structure of the paper is as follows,
\begin{itemize}
  \item Section~\ref{sec:II} introduces the ``Geometrical Triangle Hypothesis", develops its mathematical framework, and derives implications for mixing angles and mass-squared differences.
  \item Section~\ref{sec:III} explores theoretical consequences, including compatibility with global-fit values, angle adjustment under constraints, and implications for predictive structure.
  \item Section~\ref{sec:IV} outlines experimental prospects, identifying how precision measurements from current and future experiments (e.g., DUNE, Hyper-Kamiokande) could validate or falsify the triangle relation.
  \item Section~\ref{sec:V} summarizes our findings and outlines future directions in both theory and experiment.
\end{itemize}


\section{The Geometrical Triangle Hypothesis}\label{sec:II}

In the standard three-flavor neutrino oscillation framework, the flavor eigenstates $\nu_\alpha$ ($\alpha = e, \mu, \tau$) are related to the mass eigenstates $\nu_i$ ($i = 1, 2, 3$) through the unitary PMNS matrix $U$ as
\begin{equation}
\nu_\alpha = \sum_{i=1}^3 U_{\alpha i} \nu_i.
\label{eq:unitarymatrix}
\end{equation}

\begin{figure}[ht]
    \centering
    \includegraphics[width=0.8\textwidth]{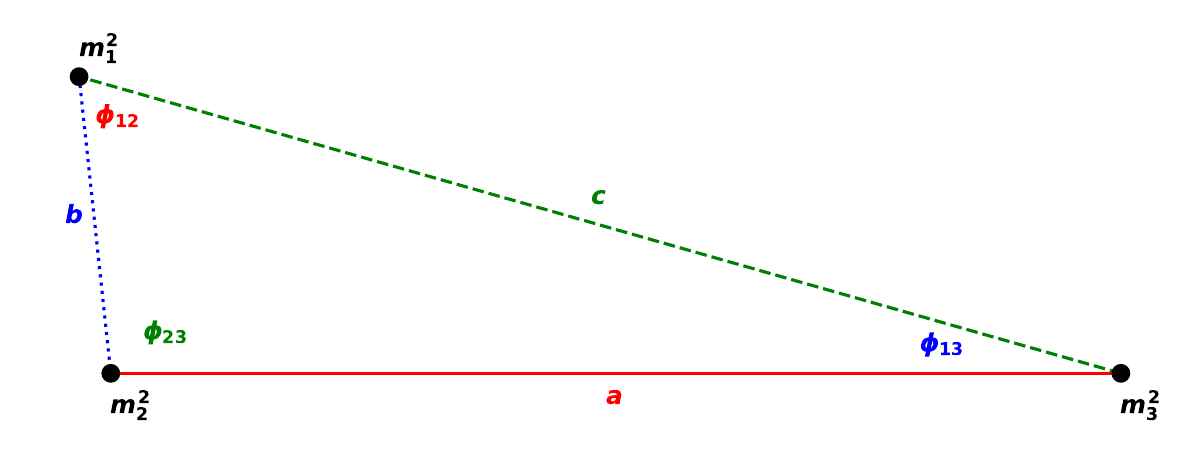}
    \caption{Mixing triangle constructed with angles $\phi_{12}, \phi_{13}, \phi_{23}$ and sides labeled $a, b, c$ corresponding to $\Delta m^2_{32}, \Delta m^2_{21}, \Delta m^2_{31}$ respectively. The vertex labels reflect mass-squared eigenvalues $m_1^2$, $m_2^2$, and $m_3^2$.}
    \label{fig:mixing_triangle}
\end{figure}

This unitary matrix is typically parametrized by three Euler mixing angles ($\theta_{12}, \theta_{13}, \theta_{23}$) and one CP-violating phase $\delta_{CP}$, with possible additional phases in the Majorana case. In this work, we explore a novel hypothesis:
\begin{quote}
    \textit{Twice the sum of the three neutrino mixing angles corresponds to the sum of the interior angles of a Euclidean triangle.}
\end{quote} 
Mathematically, we show that as Eq.~\eqref{eq:triangle}. This implies that the three angles, $\phi_{12} = 2\theta_{12}$, $\phi_{13} = 2\theta_{13}$, and $\phi_{23} = 2\theta_{23}$ may be interpreted as the interior angles of a triangle, which we shall refer to as the ``mixing triangle" as shown in Fig.~\ref{fig:mixing_triangle}.

\subsection{Motivation and Interpretation}

The motivation for Eq.~\eqref{eq:triangle} stems from the observation that mixing angles appear symmetrically in oscillation probabilities and enter into the parametrization of the PMNS matrix through rotation matrices. Although there is no direct geometric requirement for these angles to obey a Euclidean constraint, their empirical values suggest that such a relation is not far-fetched. We emphasize that this ansatz is introduced as a phenomenological organizing principle, not as a consequence of an underlying flavor symmetry. Its appeal lies in its simplicity and falsifiability. Assuming Eq.~\eqref{eq:triangle} holds, and using current best-fit values for NO from NuFIT~6.0~\cite{Esteban2024}, we calculate,
\begin{equation}
\begin{aligned}
\phi_{12} &= 2\theta_{12} = (66.82 \pm 1.47)^\circ,\\
\phi_{13} &= 2\theta_{13} = (17.08 \pm 0.23)^\circ,\\
\Rightarrow~~ \phi_{23} &= 180.00^\circ - \phi_{12} - \phi_{13} = (96.10 \pm 1.49)^\circ ~~\Rightarrow ~~\theta_{23}^{\rm pred} = \tfrac{1}{2}\phi_{23} = (48.05 \pm 0.74)^\circ.
\end{aligned}
\label{eq:angles}
\end{equation}

Here the uncertainties are propagated at $1\sigma$: $\sigma(\phi_{12})=2\sigma(\theta_{12})$, $\sigma(\phi_{13})=2\sigma(\theta_{13})$, and $\sigma(\phi_{23})=\sqrt{\sigma(\phi_{12})^2+\sigma(\phi_{13})^2}$, so, $\sigma(\theta_{23}^{\rm pred})=\frac{1}{2}\sigma(\phi_{23})$. This value of $\theta_{23}$ lies in the second octant ($\theta_{23} > 45^\circ$), which provides a direct resolution to the octant ambiguity. Moreover, the small deviation from the global best-fit, $\theta_{23} \approx 49.1^\circ$, suggests that the triangle condition provides a good approximation. 

Using Eq.~\eqref{eq:params}, we compare $\theta_{23}^{\rm pred} = 48.05^\circ$ with the fit $\theta_{23}^{\rm exp} = 49.10^\circ$ and adopt a symmetrized $1\sigma$ error $\sigma(\theta_{23}) = 1.15^\circ$ for the fit value. Then
\begin{equation}
\chi^2 = \frac{(\theta_{23}^{\text{pred}} - \theta_{23}^{\text{exp}})^2}{\sigma^2(\theta_{23})}
= \frac{(48.05^\circ - 49.10^\circ)^2}{(1.15^\circ)^2} \approx 0.83,
\end{equation}
which (for one degree of freedom) corresponds to a $p$-value of $\sim 0.36$, i.e., fully consistent within $1\sigma$. A similar one-parameter check for the angle sum uses $S = \theta_{12} + \theta_{13} + \theta_{23}$: with $S^{\rm exp} = 91.05^\circ$ and $S^{\rm pred} = 90.00^\circ$, and $\sigma(S) = \sqrt{0.75^2+0.12^2+1.15^2} \approx 1.38^\circ$, one finds $\chi^2 \approx 0.58$ ($p$-value $\sim 0.45$), again indicating compatibility.

\subsection{Constructing the Mixing Triangle}

To illustrate the geometrical hypothesis, we construct a triangle whose interior angles are given by $\phi_{ij} = 2\theta_{ij}$, using the mixing angles $\theta_{12}, \theta_{13}$, and the predicted $\theta_{23} = 48.05^\circ$. These yield the three angles of the mixing triangle as Eqs.~\eqref{eq:angles}. We associate the triangle sides opposite to these angles with the three mass-squared differences,
\begin{equation}
\begin{aligned}
  a & \leftrightarrow \Delta m^2_{32} ~~(\mathrm{opposite} ~~\phi_{12}) \\
  b & \leftrightarrow \Delta m^2_{21} ~~(\mathrm{opposite} ~~\phi_{13}) \\
  c & \leftrightarrow \Delta m^2_{31} ~~(\mathrm{opposite} ~~\phi_{23})
\end{aligned}
\label{eq:sides}
\end{equation}

This correspondence reflects the empirical size hierarchy $\Delta m^2_{21} \ll \Delta m^2_{3\ell}$ and allows us to fix $\Delta m^2_{21}$ as a scale. Using the law of sines,
\begin{equation}
\frac{a}{\sin\phi_{12}} = \frac{b}{\sin\phi_{13}} = \frac{c}{\sin\phi_{23}},
\label{eq:lawofsines}
\end{equation}
and setting $b = \Delta m^2_{21} = 7.41 \times 10^{-5}\,\mathrm{eV}^2$, we predict,
\begin{equation}
\Delta m^2_{32}{}^\mathrm{pred} = (2.32 \pm 0.05)\times 10^{-4}\,\mathrm{eV}^2, \quad
\Delta m^2_{31}{}^\mathrm{pred} = (2.51 \pm 0.05)\times 10^{-4}\,\mathrm{eV}^2.
\label{eq:masssquared_pred}
\end{equation}
The uncertainties reflect propagation from Eq.~\eqref{eq:angles} and the quoted error on $\Delta m^2_{21}$. These values are to be compared with the experimental determinations~\cite{Esteban2024, PDG2024},
\begin{equation}
\Delta m^2_{32}{}^\mathrm{exp} = (2.44 \pm 0.03)\times 10^{-3}\,\mathrm{eV}^2, \quad
\Delta m^2_{31}{}^\mathrm{exp} = (2.51 \pm 0.03)\times 10^{-3}\,\mathrm{eV}^2.
\label{eq:masssquared_exp}
\end{equation}
This result highlights that the triangle construction accurately encodes the relative proportions among the mass-squared differences, but underestimates their absolute scale by about an order of magnitude. We interpret this as evidence that the mixing triangle, defined by the angles $\phi_{ij} = 2\theta_{ij}$, captures the dimensionless geometry or hierarchy pattern of neutrino parameters, while the absolute mass scale arises from separate physics.
\begin{figure}[ht]
    \centering
    \includegraphics[width=0.99\textwidth]{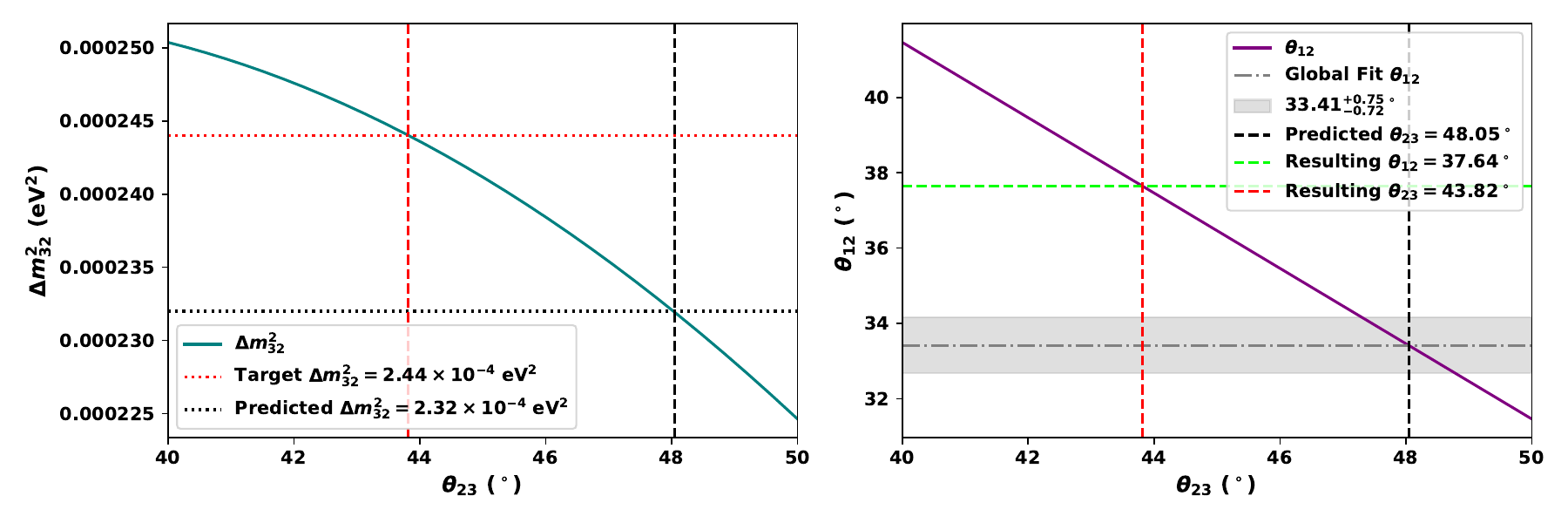}
    \caption{Variation of the triangle-predicted quantities with $\theta_{23}$. \textbf{Left:} Scaled value of $\Delta m^2_{32}$ as a function of $\theta_{23}$, assuming the triangle condition and fixing $\Delta m^2_{21}$, $\Delta m^2_{31}$ and $\theta_{13}$.  \textbf{Right:} The corresponding value of $\theta_{12}$ obtained via the triangle constraint $\theta_{12} = 90^\circ - \theta_{13} - \theta_{23}$. The shaded region shows the global-fit uncertainty range for $\theta_{12}$, which is visibly exceeded when $\theta_{23}$ is tuned to match $\Delta m^2_{32}$.}
    \label{fig:triangle_variation}
\end{figure}
We also observe, as shown in Fig.~\ref{fig:triangle_variation}, that the triangle-predicted scaled value of $\Delta m^2_{32}$ can be brought into exact agreement with experiment by decreasing $\theta_{23}$ from $48.05^\circ$ to $43.82^\circ$. However, this shift pushes the implied value of $\theta_{12}$ to $37.64^\circ$, which lies beyond the $1\sigma$ global-fit range. This highlights a mild internal tension in satisfying all mixing parameters simultaneously within the triangle hypothesis, while still supporting its predictive value.

The triangle should thus be understood as a representation of normalized mass-squared structure, potentially reflecting quantities like $\sqrt{\Delta m^2}$ or $1/m_i$ rather than $\Delta m^2_{ij}$ directly. The full physical scale, especially the atmospheric--solar mass gap, may be set by high-energy mechanisms such as the seesaw framework or radiative mass generation, where different physical origins explain the multi-scale structure. In this light, the triangle reflects internal symmetry or flavor structure, while the overall mass-squared scale is determined externally.

\subsection{General Framework and Mass Ordering Interpretation}
The triangle hypothesis proposes that the three mixing angles \(\theta_{12}, \theta_{13}, \theta_{23}\) define a Euclidean triangle via their doubled values \(\phi_{ij} = 2\theta_{ij}\). This triangle, fixed in shape by experimental data, serves as a geometric scaffold for interpreting neutrino mass structure. The triangle’s vertices are labeled by the squared mass eigenvalues \(m_1^2\), \(m_2^2\), and \(m_3^2\), while its sides correspond to the absolute values of mass-squared differences \(\Delta m^2_{ij} = |m_i^2 - m_j^2|\). A key feature of this framework is that the triangle does not encode the mass ordering explicitly. Instead, the ordering is inferred by choosing a starting vertex and direction of traversal (clockwise or anti-clockwise), and interpreting only two sides as mass-squared differences. The third side merely closes the triangle and is not used for inference.

\noindent\textbf{Consistent Mass Orderings from Anti-Clockwise Traversals:}
\begin{itemize}
    \item \textbf{Normal Ordering:} Start at vertex \(m_3^2\), traverse to \(m_1^2 \rightarrow m_2^2\)
    \[
    10\times c = |m_3^2 - m_1^2| \Rightarrow m_3 > m_1,\quad
    b = |m_1^2 - m_2^2| \Rightarrow m_1 > m_2 \Rightarrow m_3 > m_1 > m_2.
    \]
    
    \item \textbf{Inverted Ordering:} Start at vertex \(m_2^2\), traverse to \(m_3^2 \rightarrow m_1^2\)
    \[
    10\times a = |m_2^2 - m_3^2| \Rightarrow m_2 > m_3,\quad
    10\times c = |m_3^2 - m_1^2| \Rightarrow m_3 > m_1 \Rightarrow m_2 > m_3 > m_1.
    \]
    
    \item \textbf{Mixed Ordering:} Start at vertex \(m_1^2\), traverse to \(m_2^2 \rightarrow m_3^2\)
    \[
    b = |m_1^2 - m_2^2| \Rightarrow m_1 > m_2,\quad
    10\times a = |m_2^2 - m_3^2| \Rightarrow m_2 > m_3 \Rightarrow m_1 > m_2 > m_3.
    \]
\end{itemize}

\noindent\textbf{Consistent Mass Orderings from Clockwise Traversals:}
\begin{itemize}
    \item \textbf{Normal Ordering:} Start at vertex \(m_1^2\), traverse to \(m_3^2 \rightarrow m_2^2\)
    \[
    10\times c = |m_3^2 - m_1^2| \Rightarrow m_3 > m_1,\quad
    b = |m_2^2 - m_1^2| \Rightarrow m_1 > m_2 \Rightarrow m_3 > m_1 > m_2.
    \] 
    \item \textbf{Inverted Ordering:} Start at vertex \(m_3^2\), traverse to \(m_2^2 \rightarrow m_1^2\)
    \[
    10\times a = |m_2^2 - m_3^2| \Rightarrow m_2 > m_3,\quad
    10\times c = |m_1^2 - m_3^2| \Rightarrow m_3 > m_1 \Rightarrow m_2 > m_3 > m_1.
    \]
    \item \textbf{Mixed Ordering:} Start at vertex \(m_2^2\), traverse to \(m_1^2 \rightarrow m_3^2\)
    \[
    b = |m_2^2 - m_1^2| \Rightarrow m_2 > m_1,\quad
    10\times a = |m_1^2 - m_3^2| \Rightarrow m_1 > m_3 \Rightarrow m_2 > m_1 > m_3.
    \]
\end{itemize}

Each of these cases results from interpreting only two adjacent sides from a traversal path, with each path yielding a distinct, transitive mass hierarchy. A striking feature of the triangle construction is that starting from the same vertex, different traversal directions yield different mass orderings. For instance, beginning at \(m_3^2\) and moving anti-clockwise leads to the normal ordering \(m_3 > m_1 > m_2\), while starting at \(m_3^2\) and traversing clockwise instead leads to the inverted ordering \(m_2 > m_3 > m_1\). This illustrates that the mass hierarchy is not solely determined by vertex labels, but is intrinsically tied to the direction of traversal, reinforcing the idea that the triangle encodes multiple possible orderings in a direction-dependent way. Note: these `mixed' orderings ($m_1 > m_2 > m_3$ from anti-clockwise and $m_2 > m_1 > m_3$ from clockwise traversals) are effectively the normal or inverted ordering if one relabels the states by mass. We include it for completeness as a distinct traversal outcome of the triangle, though physically the neutrino mass spectrum is usually categorized only as normal or inverted.

\noindent\textbf{Inconsistency of Cyclic Traversals:}

Attempting to interpret all three sides by completing a full loop around the triangle (in either direction) leads to a contradiction. For example, a clockwise traversal from \(m_3^2 \rightarrow m_2^2 \rightarrow m_1^2 \rightarrow m_3^2\) implies
\[
m_3 > m_2,\quad m_2 > m_1,\quad m_1 > m_3,
\]
which is cyclic and inconsistent. This reinforces that only two edges should be interpreted per mass ordering inference, a geometric reflection of the fact that only two independent mass-squared differences exist in three-flavor oscillations.


\section{Theoretical Implications and Predictive Power}\label{sec:III}

The triangle hypothesis introduces a geometric constraint on neutrino mixing angles that departs from traditional model-building approaches. Rather than deriving mixing patterns from flavor symmetries or grand unified frameworks, it postulates a structural relation among the mixing angles themselves: their doubled values form the interior angles of a Euclidean triangle. This section examines how such a hypothesis aligns or conflicts with existing theories, what new insights it offers, and how it might be interpreted in a deeper physical context.

\subsection{Quantitative Predictivity}
Eq.~\eqref{eq:angles} yields $\theta_{23}^{\rm pred} = (48.05 \pm 0.74)^\circ$,
to be compared with the global fit $\theta_{23}^{\rm exp} = (49.10^{+1.00}_{-1.30})^\circ$. The $1.05^\circ$ offset corresponds to $\chi^2 \approx 0.83$ (Sec.~\ref{sec:II}), indicating agreement within $1\sigma$. Notably, the uncertainty on $\theta_{23}^{\rm pred}$ is smaller than that of the global fit, showing that the triangle relation effectively reduces one degree of freedom. Thus, as data improve, the hypothesis will either be sharply confirmed or excluded.

\subsection{Comparison with Symmetry-Based Models}

Many theoretical approaches derive specific mixing patterns from discrete symmetries. For example, TBM mixing predicts
\[
\theta_{12} = 35.26^\circ,\quad \theta_{23} = 45.00^\circ,\quad \theta_{13} = 0.00^\circ~\cite{Harrison2002},
\]
so that
\[
2(\theta_{12} + \theta_{23} + \theta_{13}) = 160.52^\circ,
\]
well below the $180^\circ$ triangle condition. BM mixing gives
\[
\theta_{12} = 45.00^\circ,\quad \theta_{23} = 45.00^\circ,\quad \theta_{13} = 0.00^\circ,
\]
which yields $2(\sum \theta) = 180.00^\circ$ exactly, but disagrees strongly with experiment since $\theta_{12}$ is $\sim 12^\circ$ too large and $\theta_{13}=0$. Golden Ratio (GR) mixing predicts $\theta_{12} \approx 31.7^\circ$ or $36.0^\circ$ depending on the variant~\cite{Everett2009}, and mu–tau symmetry~\cite{Mohapatra2005} also fails to enforce the $90^\circ$ sum. Neither approach matches the triangle relation. Thus, BM is numerically consistent with the sum rule but phenomenologically excluded, TBM and GR approximate observed angles but fail the sum rule, while the triangle ansatz achieves both: it satisfies the $90^\circ$ condition and matches current best-fit values within $\sim 1^\circ$.

\subsection{Mass-Splitting Ratios}

Using the law of sines construction, we found Eq.~\eqref{eq:masssquared_pred}
to be compared with Eq.~\eqref{eq:masssquared_exp}. Numerically, the experimental values are about $10\times$ larger than the triangle predictions, but the ratios are consistent. Defining
\begin{equation}
\begin{aligned}
\alpha^{\rm pred} &= \frac{\Delta m^2_{21}{}^{\rm exp}}{\Delta m^2_{31}{}^{\rm pred}} \approx 0.030,
\quad
\alpha^{\rm exp} = \frac{\Delta m^2_{21}{}^{\rm exp}}{\Delta m^2_{31}{}^{\rm exp}} \approx 0.029,\\
\beta^{\rm pred} &= \frac{\Delta m^2_{32}{}^{\rm pred}}{\Delta m^2_{31}{}^{\rm pred}} \approx 0.92,
\quad
\beta^{\rm exp} = \frac{\Delta m^2_{32}{}^{\rm exp}}{\Delta m^2_{31}{}^{\rm exp}} \approx 0.97.
\end{aligned}
\label{eq:ratios}
\end{equation}
we see that the triangle reproduces $\alpha^{\rm pred}$ and $\beta^{\rm pred}$ within a few percent of the global-fit values, even though it underestimates the absolute splittings by an order of magnitude. This strongly suggests that the triangle encodes the dimensionless structure of the mass hierarchy while the absolute scale arises from separate physics, such as seesaw mass generation~\cite{Minkowski1977}. The $\sim 10\times$ scale mismatch may reflect a physical separation between the solar and atmospheric mass-generation sectors, a feature that appears in many seesaw-based frameworks. Thus, the triangle does not replace detailed model-building, but may serve as a compact phenomenological summary of the low-energy consequences of those models.

\subsection{Speculative Extensions: CP Phase $\delta_{CP}$}
The present triangle hypothesis does not constrain $\delta_{CP}$, yet this parameter is central to the discovery reach of DUNE and Hyper-Kamiokande. By analogy with the CKM unitarity triangle, $\delta_{CP}$ may be interpreted geometrically. The leptonic Jarlskog invariant,
\begin{equation}
J_\nu = \tfrac{1}{8}\sin 2\theta_{12}\,\sin 2\theta_{13}\,\sin 2\theta_{23}\,\sin \delta_{CP},
\end{equation}
is proportional to the area of the leptonic unitarity triangle. One could envisage extending the mixing triangle into the complex plane, where $\delta_{CP}$ appears as an orientation or area factor ensuring closure. Many flavor models predict $\delta_{CP}$ near $90^\circ$ or $270^\circ$, coinciding with large CP violation. Since the triangle ansatz prefers $\theta_{23}$ in the higher octant ($\sim 48^\circ$), it is natural to expect large $\delta_{CP}$ as well. While speculative, this extension connects our framework directly to the upcoming experimental program, and provides an avenue for future theoretical development. We stress that this possible extension to $\delta_{CP}$ is exploratory and qualitative, intended only as a suggestion for future work rather than a firm prediction.


\section{Experimental Prospects and Tests}\label{sec:IV}

The triangle hypothesis makes specific, testable predictions that can be confronted with increasingly precise neutrino data. While it introduces no new particles or interactions, it tightly correlates the three mixing angles and indirectly constrains the neutrino mass-squared differences. This makes it accessible to near-future precision oscillation experiments.

\begin{itemize}
    \item \textbf{Angle correlation test:} The core condition in Eq.~\eqref{eq:triangle} is not guaranteed by the Standard Model or any known symmetry. Precise measurements of all three angles can directly confirm or refute this relation. As of current global fits, the triangle condition holds within $\sim 1^\circ$. Future tests will require uncertainties on $\theta_{12}, \theta_{13}, \theta_{23}$ at the level of $0.2^\circ$--$0.3^\circ$ to discriminate decisively.

    \item \textbf{Octant prediction:} The triangle naturally favors a second-octant value of $\theta_{23}$, predicting $\theta_{23}^{\rm pred} = (48.05 \pm 0.74)^\circ$. If experiments establish $\theta_{23}<45^\circ$ with $\gtrsim 3\sigma$ significance, the hypothesis would be excluded. Conversely, a confirmed high-octant $\theta_{23}$ with uncertainty $\lesssim 0.3^\circ$ would provide a stringent test of the predicted value.

    \item \textbf{Mass-squared ratio matching:} Given $\theta_{12}$ and $\theta_{13}$, the triangle predicts $\theta_{23}$ and, via the law of sines, the ratios $\alpha^{\rm pred}$ and $\beta^{\rm pred}$ in Eq.~\eqref{eq:ratios}. The current agreement of $\alpha^{\rm pred}=0.030$ and $\beta^{\rm pred}=0.92$ with experimental $\alpha^{\rm exp}\approx 0.029$ and $\beta^{\rm exp}\approx 0.97$ shows that ratios are reproduced to within a few percent. Improved measurements of $\Delta m^2_{21}$ at JUNO (target precision $\lesssim 0.5\%$)~\cite{JUNO2016,Abusleme2022} and of $\Delta m^2_{31}$ at DUNE/Hyper-K (target precision $\sim 1\%$)~\cite{Abi2020,Kose2025} will test these relations precisely.

    \item \textbf{Future experimental sensitivity:} DUNE~\cite{Abi2020} and Hyper-Kamiokande~\cite{Kose2025} aim to measure $\theta_{23}$ to $\pm 0.3^\circ$ and $\delta_{CP}$ to $\pm 10^\circ$--$15^\circ$, while JUNO~\cite{JUNO2016,Abusleme2022} will reach $\sim 0.2^\circ$ precision on $\theta_{12}$. These precisions are sufficient to decisively confirm or reject the $90^\circ$ sum rule. A deviation at the $\gtrsim 3\sigma$ level would falsify the hypothesis.

    \item \textbf{Extensions to CP phase:} If the triangle hypothesis is extended to include $\delta_{CP}$, it may imply geometric constraints related to the area or orientation of a complexified triangle. A non-trivial prediction is that large CP violation (e.g., $\delta_{CP}\approx 90^\circ$ or $270^\circ$) should accompany the preferred higher-octant $\theta_{23}$. Ongoing measurements at T2K, NO$\nu$A, and in the future DUNE and Hyper-K will be able to test this feature directly.
\end{itemize}

In summary, the triangle hypothesis offers clear falsification criteria:
\begin{enumerate}
    \item If $\theta_{23}$ is confirmed in the first octant ($<45^\circ$), the framework is ruled out.
    \item If $\theta_{12}+\theta_{13}+\theta_{23}$ deviates from $90^\circ$ by more than $\sim 1^\circ$ once the errors shrink below $0.3^\circ$, the hypothesis is excluded.
    \item If the ratios $\alpha$ or $\beta$ deviate from their predicted values by more than a few percent, the geometric consistency breaks down.
\end{enumerate}

The triangle hypothesis thus serves as a predictive and falsifiable organizing principle. Even if it is eventually shown to be only approximate, its success in capturing known features of neutrino mixing makes it a valuable benchmark for theoretical exploration.


\section{Conclusion and Outlook}\label{sec:V}

We have investigated a phenomenological ``triangle hypothesis'' for neutrino mixing, in which twice the sum of the three mixing angles forms the interior angles of a Euclidean triangle. Using NuFIT~6.0 data, we showed that this relation predicts $\theta_{23}^{\rm pred} = (48.05 \pm 0.74)^\circ$, consistent within $1\sigma$ of the global-fit value, and reproduces the observed ratios of mass-squared differences at the few-percent level. A simple $\chi^2$ check confirms present compatibility.

The framework is distinctive because it reduces one free parameter among the mixing angles, thereby making falsifiable predictions: the sum $\theta_{12}+\theta_{13}+\theta_{23}$ must equal $90^\circ$, and $\theta_{23}$ is constrained to the higher octant near $48^\circ$. The observed $\sim 10\times$ mismatch in absolute mass-squared scales suggests that the triangle encodes dimensionless structure, while the overall scale originates in separate physics such as the seesaw.

Future experiments will provide decisive tests. JUNO will sharpen $\theta_{12}$ and $\Delta m^2_{21}$ to sub-percent precision, while DUNE and Hyper-Kamiokande will reduce errors on $\theta_{23}$ and $\delta_{CP}$. These measurements will either validate the triangle relation as a useful organizing principle or exclude it at high significance.

In either outcome, the exercise of casting the PMNS parameters into a simple geometric structure enriches the search for patterns in lepton flavor. If confirmed, the triangle hypothesis could point to deeper symmetries or alignments; if excluded, it will still serve as a benchmark for understanding why neutrino mixing exhibits its strikingly non-hierarchical form.

\begin{acknowledgments}
This work was supported in part by the National Science Foundation (NSF) under Grants No.~2310096, 2310126, and 2012989, and by the Ralston Dissertation Fellowship, Department of Physics \& Astronomy, University of Kansas. 
\end{acknowledgments}


\section*{Author Declarations}

\noindent\textbf{Conflict of Interest:} The author declares no conflicts of interest.\\[0.5em]
\noindent\textbf{Data Availability:} This study uses publicly available global-fit neutrino oscillation data, as cited in the references. Additional numerical data generated by the author for figure construction are available from the author upon reasonable request.\\[0.5em]
\noindent\textbf{Author Contributions:} The author is solely responsible for all aspects of the research and manuscript preparation.

\bibliographystyle{aipnum4-2}
\renewcommand{\bibfont}{\small}
\setlength{\bibsep}{1.0pt}     
\bibliography{references}

\end{document}